\documentclass[journal]{IEEEtran}

\usepackage{subcaption}
\usepackage{amssymb}
\usepackage{amsfonts}
\usepackage{amsthm}
\usepackage{amsmath}
\usepackage{cite}
\usepackage[]{graphicx}
\usepackage{caption}
\usepackage{algorithmic}
\usepackage{eufrak}

\usepackage[hyphens]{url}
\usepackage[hidelinks]{hyperref}
\hypersetup{breaklinks=true}
\usepackage[english]{babel}
\usepackage{array}
\newcolumntype{L}{>{\raggedright\arraybackslash}m{4.7cm}}
\usepackage{color}
\usepackage{graphicx}
\usepackage[export]{adjustbox}
\usepackage{amsmath}
\usepackage{cleveref}
\usepackage{mathtools}

\usepackage[ruled]{algorithm2e}

\usepackage{lipsum}
\usepackage[shortlabels]{enumitem}

\theoremstyle{definition}
\newtheorem{definition}{Definition}[section]

\begin{document}

\title{Achieving Correlated Equilibrium by Studying Opponent's Behavior Through Policy-Based Deep Reinforcement Learning}

\author{Kuo~Chun~Tsai,~\IEEEmembership{Student Member,~IEEE}
        and~Zhu~Han,~\IEEEmembership{Fellow,~IEEE}

\IEEEcompsocitemizethanks{
\IEEEcompsocthanksitem Kuo Chun Tsai is with the Department of Electrical and Computer Engineering, University of Houston, Houston, TX, USA. E-mail: kevintsai159@gmail.com.
\IEEEcompsocthanksitem Zhu Han is with the Department of Electrical and Computer Engineering, University of Houston, Houston, TX, USA, and also with the Department of Computer Science and Engineering, Kyung Hee University, Seoul, South Korea. E-mail: zhan2@uh.edu.}
}

\maketitle

\begin{abstract}
Game theory is a very profound study on distributed decision-making behavior and has been extensively developed by many scholars. However, many existing works rely on certain strict assumptions such as knowing the opponent's private behaviors, which might not be practical. In this work, we focused on two Nobel winning concepts, the Nash equilibrium and the correlated equilibrium. Specifically, we successfully reached the correlated equilibrium outside the convex hull of the Nash equilibria with our proposed deep reinforcement learning algorithm. With the correlated equilibrium probability distribution, we also propose a mathematical model to inverse the calculation of the correlated equilibrium probability distribution to estimate the opponent's payoff vector. With those payoffs, deep reinforcement learning learns why and how the rational opponent plays, instead of just learning the regions for corresponding strategies and actions. Through simulations, we showed that our proposed method can achieve the optimal correlated equilibrium and outside the convex hull of the Nash equilibrium with limited interaction among players.
\end{abstract}


%
\IEEEpeerreviewmaketitle

\section{Introduction}\label{introduction}
Game theory is the study of mathematical models regarding the rationality decision making strategic interaction. Originally, game theory addressed the two-player zero-sum non-cooperative games where each participant's gains or losses are exactly balanced by those of the other participants \cite{bowles2009microeconomics}. Today, game theory applies to a wide range of behavioral relations such as cooperative game \cite{branzei2008models}, contract theory, auction theory, matching game, dynamic game and more\cite{han2011game,han_niyato_saad_basar_2019}.

The main idea for game theory is for each rational player to maximize his or her utility. Take the Nobel winning Nash equilibrium solution concept for example. The Nash equilibrium solution concept, named after the mathematician John Forbes Nash Jr., is a solution of a non-cooperative game involving two or more players. A Nash equilibrium in the game is where no player has utility increment by changing only their strategy \cite{osborne1994course}. However, the solution under the Nash equilibrium might be far from the centralized solution. Hence, in 1974, there is another mathematician named Robert Aumann discussed another Nobel winning solution concept called correlated equilibrium \cite{AUMANN197467}. The idea in the correlated equilibrium concept is that all players will choose an action according to a public signal. If all player is satisfied by the recommended strategy, the distribution is called a correlated equilibrium distribution \cite{fudenberg1991game}. With this concept, players can achieve higher utility in the game since they consider the joint distribution instead of the marginal distribution as the Nash equilibrium. However, there does not exist a distributed strategy on how to reach correlated equilibrium outside the Nash equilibrium convex hull since there does not exist a clear way on how to design the public signal for the player to obtain. Some existing works using non-regret learning \cite{4224253} can only achieve the Nash equilibrium convex hull but cannot learn the better correlated equilibrium.

The reason that the players in a non-cooperative game are unable to achieve the correlated equilibrium is due to the fact that they are unable to mine some of the information out from the public signal provided to them. Fortunately, a technique called machine learning was developed for studying the underlying factor of the data. Many works, such as \cite{10_1007_3_540_70659_3_2, 8353149, 8880673, 6423821} and more, showed the robustness of the machine learning technique for data mining tasks. However, even today, the link of machine learning with game theory is seldom studied.

Motivated by the above facts, we proposed a policy-based deep reinforcement learning model to determine the strategy to reach equilibrium under limited information given to the player. We then estimate the payoffs of other players via our proposed mathematical model. Once we have all the payoffs from the other players, we can determine the correlated equilibrium between other players without them interacting with each other. Thus, the contribution we made in this work can be summarized as follows:

\begin{itemize}
\item We first define the public signal in the system for the player to obtain that contains limited information of the players.

\item With limited information, players will start to interact with the environment. By applying our proposed deep reinforcement learning model, the player not only can understand the structure of the environment but also learn the joint distribution among all of the players when exploring the environment. In the end, the players can reach a correlated equilibrium where no one wants to deviate.

\item With the correlated equilibrium probability distribution that all players satisfied with, we proposed a mathematical model that combines the concept of the correlated equilibrium and the force of tension to estimate the payoff vectors of other players. 

\item By knowing the payoff vector of other players in the system, we could compute the correlated equilibrium. In other words, with those payoff vectors, deep reinforcement learning learns why and how the rational opponent plays, instead of just learning the regions for corresponding strategies and actions.

\item This paper combines the game theory with machine learning in the sense that the proposed machine learning learns what is the game player's payoff, instead of just categorizing the strategies of actions according to the current situation.
\end{itemize}

The rest of the paper is organized as follows. In Section \ref{system_model_and_basic_equilibrium_concept}, we discuss the basic concept in the Nash equilibrium and correlated equilibrium concepts along with our system model and problem formulation. Next, in Section \ref{proposed_method}, we show how the player interacts with the environment with our proposed policy-based deep reinforcement learning model and learn the joint distribution among the players. In the same section, we also proposed a mathematical model to estimate the information of the opponent players. Next, we show the numerical results for our proposed methods in Section \ref{performance_evaluation}. Finally, we conclude our work in Section \ref{conclusion}.

\section{System Model and Basic Equilibrium Concept}\label{system_model_and_basic_equilibrium_concept}
In this section, we will first go through the basic concept of the Nash equilibrium and the correlated equilibrium in Section \ref{nash_equilibrium_and_correlated_equilibrium}. In Section \ref{system_model}, we will study the relationship between the environment and the players in our system model. Finally, in Section \ref{problem_formulation}, we will discuss the problem formulation.

\subsection{Nash Equilibrium and Correlated Equilibrium Basics}\label{nash_equilibrium_and_correlated_equilibrium}
The Nash equilibrium is a solution concept for a non-cooperative game for two or more players \cite{osborne1994course}. The equilibrium outcome of a non-cooperative game is one where no player wants to deviate from his or her chosen strategy after considering the opponent's decision. In other words, an individual cannot increment his or her utility from unilaterally changing his or her strategies, assuming the other players remain their strategies. There might be none or multiple Nash equilibrium in a non-cooperative game depends on the setup of the game and the strategies used by each player. A formal definition of Nash equilibrium is as follows.

\theoremstyle{definition}
\begin{definition}\label{ne_def_all}
An $I$-player game is characterized by an action set ${\Phi}_i$. Let ${B_i}\left( {{\varsigma}_{-i}} \right)\subset {\Phi}_i$ be the set of player $p_i$'s best response strategy against ${\varsigma}_{-i} \in {\Phi}_{-1}$. ${{\varsigma}^*}=\left( {{\varsigma}_{1}^*,\dots,{\varsigma}_{I}^*} \right) \in {\Phi}$ is a Nash equilibrium if ${{\varsigma}^*} \in {B_i}\left( {{\varsigma}_{-i}^*} \right)$ for every $i \in I$.
\end{definition}

On the other hand, the correlated equilibrium is also a solution concept that is more general than the Nash equilibrium. The idea of the correlated equilibrium solution concept is that each player chooses their decision according to their observation of a public signal \cite{ 00129682}. A strategy assigns an action to every possible decision set $D_h$ a player can choose. If no player wants to deviate from the recommended strategy, the distribution is called a correlated equilibrium. A formal definition is as follows.

\theoremstyle{definition}
\begin{definition}\label{ce_def_all}
An $I$-player strategic game $\left ( I , {\Phi}_i , u_i \right )$ is characterized by an action set ${\Phi}_i$ and utility function $u_i$ for each player $i$. when player $i$ chooses strategy ${\varsigma}_i \in {\Phi}_i$ and the remaining players choose a strategy profile ${\varsigma}_{-i}$ described by the $I-1$ tuple. Then the player $i$'s utility is ${u_i} \left ( {\varsigma}_i, {\varsigma}_{-i} \right )$. A strategy modification for player $i$ is a function ${\phi _i}:{{\Phi}_i} \to {{\Phi}_i}$. That is ${\phi _i}$ tells player $i$ to modify his or her behavior by playing action ${\phi _i}\left( {\varsigma}_i \right)$ when instructed by play ${\varsigma}_i$. Let $\left( {\Omega ,\Psi } \right)$ be a countable probability space. For each player $i$, let $F_i$ be his or her information partition, $q_i$ be $i$'s posterior and let ${s_i}:\Omega  \to {{\Phi}_i}$, assigning the same value to states in the same cell of $i$'s information partition. Then $\left( {\left( {\Omega ,\Psi } \right),{F_i},{s_i}} \right)$ is the correlated equilibrium of the strategic game $\left ( I , {\Phi}_i , u_i \right )$ if it satisfies the condition

\begin{equation}\label{ce_def}
\begin{array}{l}
\sum\limits_{\omega  \in \Omega } {{q_i}\left( \omega  \right)} {u_i}\left( {{s_i}\left( \omega  \right),{s_{ - i}}\left( \omega  \right)} \right) \ge\\
\sum\limits_{\omega  \in \Omega } {{q_i}\left( \omega  \right)} {u_i}\left( {{\phi _i}\left( {{s_i}\left( \omega  \right)} \right),{s_{ - i}}\left( \omega  \right)} \right)
\end{array}
\end{equation}
for every player $i$ and for every strategy modification ${\phi _i}$.
\end{definition}

\subsection{System Model}\label{system_model}

\begin{figure}[t]
    \centering
    \includegraphics[width=0.8\columnwidth]{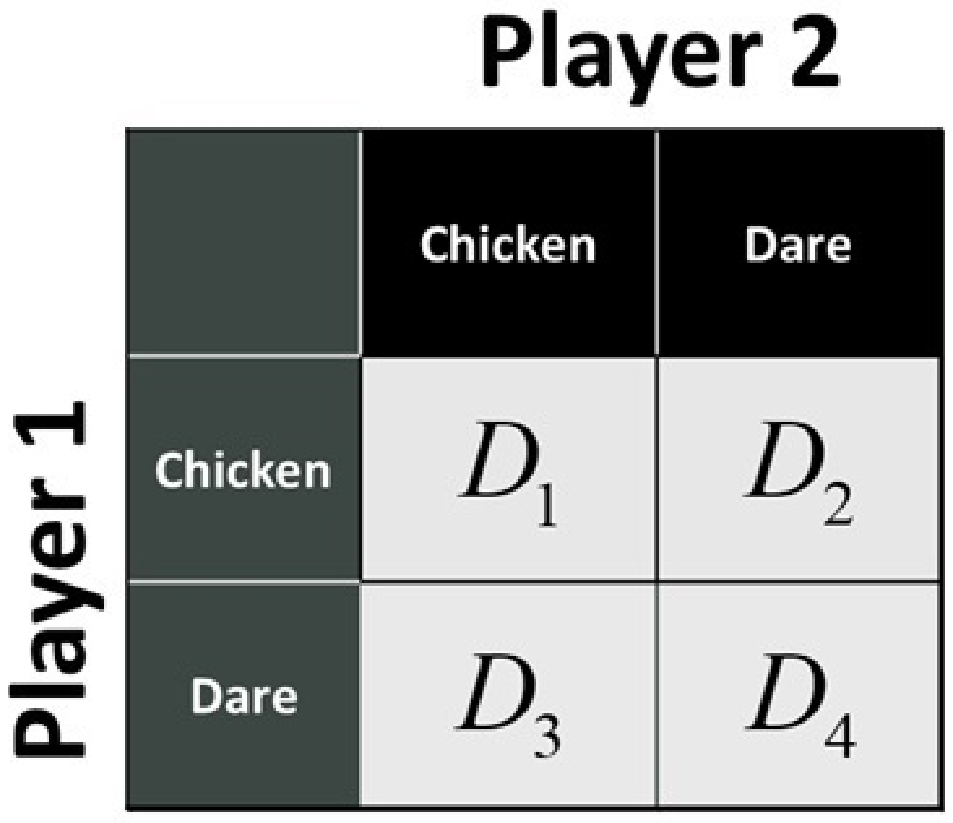}
    \caption{Game of Chicken Decision Set Layout} 
    \label{fig:chicken_game_decision_set}
\end{figure}

There are two major components in our system model – the players and the environment. We can consider the environment is a set of states $\mathbb{S}$ where each state $s_k=\left({\rho}_1,{\rho}_2,\dots,{\rho}_H\right)$ is a unique tuple which contains the probabilities for each decision set $D_h \in \mathbb{D}$ for $h=1,2,\dots,H$, where $\mathbb{D}$ is a set of permutation of all player's decisions. In other words, state $s_k$ contains a probability distribution of the decisions in $\mathbb{D}$. The order of the players' decision in decision set $D_h$ will always starts from player $p_1$ to player $p_I$. Take the game of chicken (Fig. \ref{fig:chicken_game_decision_set}) for example. The set $\mathbb{D}$ contains total four elements where $D_1$, $D_2$, $D_3$, and $D_4$ are set to (Chicken, Chicken), (Chicken, Dare), (Dare, Chicken), and (Dare, Dare), respectively.

On the other hand, there is a set of players $P$ in our system model where $P$ contains a total of $I$ players where $2 \le I \in \mathbb{Z}$. Each player $p_i$ has their own payoff vector $V_i$, policy ${\pi}_i$, and a set of actions $\mathbb{A}$. The payoff vector $V_i$ contains rewards $v_{i,h}$ that player $p_i$ will receive when agreeing on performing the decision set $D_h$ in the game. The sum of all the elements in payoff vector $V_i$ has to be equal to one. This allows us to determine the ratio between each element when we estimate the payoff vector for the other players in Section \ref{opponent_payoff_estimation}. Next, the policy ${\pi}_i$ is the behavior on how player $p_i$ will interact with the environment. Policy ${\pi}_i$ states that player $p_i$ will follow a certain probability distribution to choose an action $a_j \in \mathbb{A}$ to perform based on the current state $s_k$ that the player is currently in and the previous state $s_{k-1}$ that the player came from. The action set $\mathbb{A}$ is the permutation of increase, decrease, or no change on the probability of each of the decision set $D_h$ for $1 \le h \le H-1$. The amount of increasing or decreasing the probability is set to be $\vartheta \in \left( 0,1 \right]$. The reason why the action does not contain the change of decision $D_H$ is due to the fact that the probability distribution of all decisions in $\mathbb{D}$ needs to sum up to one. Hence, we can get the probability of $D_H$ by subtracting the summation of all other decisions' probabilities from one. In addition, even though the probability adjustment amount $\vartheta$ is a constant, if the probability is out of the range of $\left[ {0,1} \right]$ after increasing or decreasing the amount of $\vartheta$, depending on the action, the probability will only increase or decrease a certain amount so the probability will still within the range of $\left[ {0,1} \right]$. Next, depends on the action the player chooses, the player will be moving to another state. The detail on how the player will interact with the environment will be discussed in Section \ref{data_collection}.


\begin{figure}[t]
   \centering
   \includegraphics[width=0.85\columnwidth]{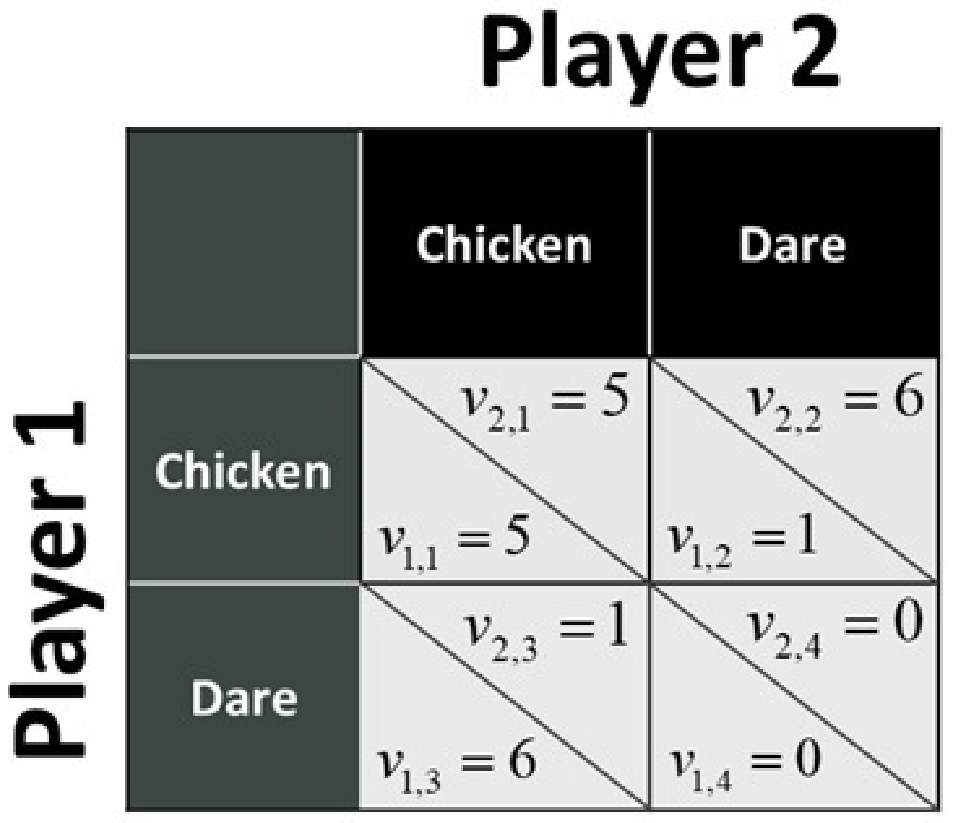}
   \caption{Game of Chicken Variables Layout} 
   \label{fig:chicken_game_variable}
\end{figure}

Moreover, according to Robert J. Aumann, the correlated equilibrium is a general form of strategy randomization than mixing. This means that the solution of the Nash equilibrium is within the convex haul of the correlated equilibrium. Hence, we set a restriction where there must be at least two Nash equilibrium sets in the system. Moreover, the Nash equilibrium sets have to be found using the mixed strategy. This restriction future express as the rewards in $V_i$ for player $p_i$ will be unique values when the decisions are fixed for other players. Otherwise, there will be no solution to the mixed strategy. Take Fig. \ref{fig:chicken_game_variable} as an example. The rewards constraint is that $v_{1,1} \ne v_{1,3}$ and $v_{1,2} \ne v_{1,4}$ for player 1, and  $v_{2,1} \ne v_{2,2}$ and $v_{2,3} \ne v_{2,4}$ for player 2.

\subsection{Problem Formulation}\label{problem_formulation}
\begin{figure}[t]
   \centering
   \includegraphics[width=0.75\columnwidth]{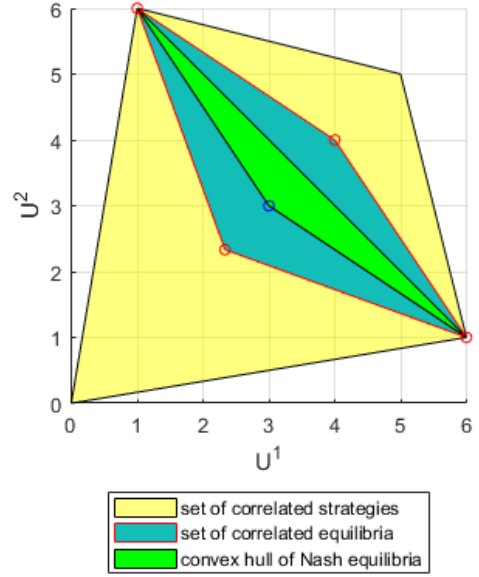}
   \caption{Nash Equilibrium vs. Correlated Equilibrium} 
   \label{fig:chicken_game_cene_convex_hull}
\end{figure}

Although players could achieve correlated equilibrium to obtain a higher reward by using the correlated strategy instead of the mixed strategy in a non-cooperative game, there exist few challenges within the correlated equilibrium solution concept itself. First, as mentioned in Section \ref{nash_equilibrium_and_correlated_equilibrium}, the correlated equilibrium concept is a more general form of strategy randomization than the mixing Nash equilibrium strategy. We can see the relation between the correlated equilibrium convex haul and the Nash equilibrium convex hull from the game of chicken shows in Fig. \ref{fig:chicken_game_cene_convex_hull}. Although it is true that the reward obtained by the player might be higher when applying the correlated strategy rather than the Nash equilibrium strategy. However, at the same time, this also means there is a probability that the player can obtain a lower reward. Second, the correlated equilibrium concept is built on the idea of having a public signal for players to observe. Based on this public signal, the player will choose the suggested decision. However, this public signal must fulfill constrain (\ref{ce_def}). This means that the public signal must be generated based on the payoff vectors from all of the players. This further implies that the payoff vectors from all of the players are public information which violates the purpose of a game. Otherwise, this will become a centralized system with centralized solutions. Hence, without knowing the payoff vectors from other players, there is no guarantee the game can reach a correlated equilibrium. Thus, the above challenges within the correlated equilibrium solution concept motivate us to design a public signal that contains limited information about each of the players and a strategy for players that guarantee them to achieve the correlated equilibrium. Moreover, the correlated equilibrium they achieved will also maximize each players' reward.

\begin{figure*}[t]
   \centering
   \includegraphics[width=\textwidth, height=6.5cm]{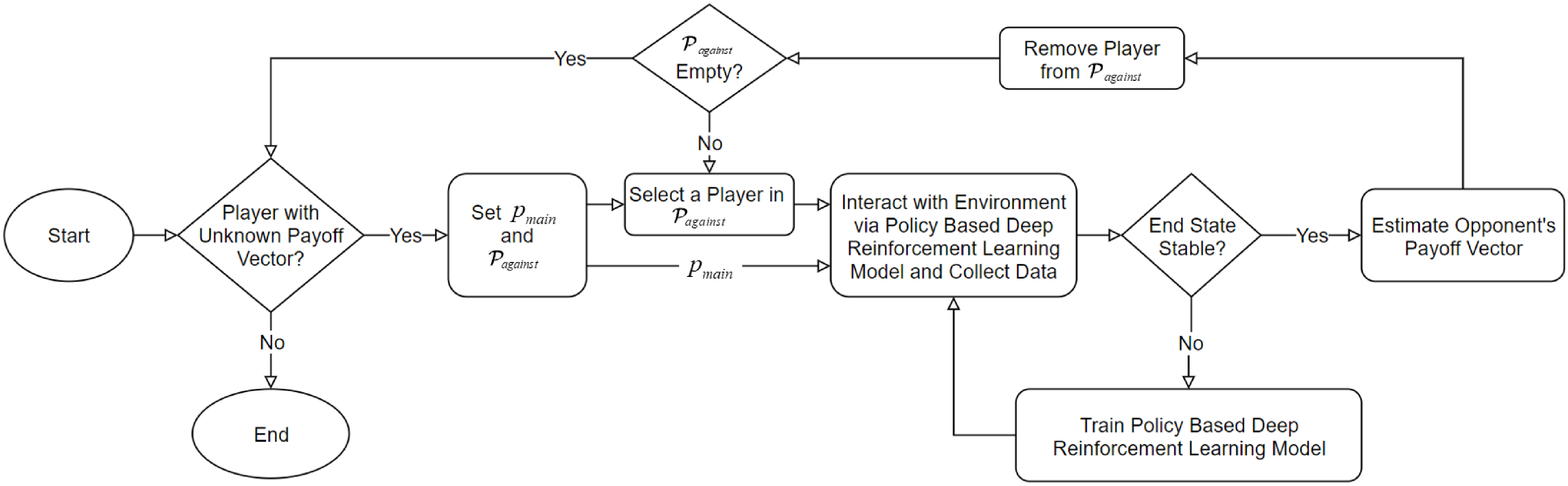}
   \caption{Proposed Method Process Flowchart} 
   \label{fig:process_flowchart}
\end{figure*}

\section{Proposed Method}\label{proposed_method}

Our proposed method will be split into two major parts – correlated equilibrium distribution determination and payoff vectors estimation. The first part is to determine the correlated equilibrium distribution between players and maximize the summation of each of the player's reward under this distribution. We first provide an overview of the proposed process in Section \ref{process_overview}. Then we discuss how players interact with the environment and collect data during the interaction in Section \ref{data_collection}. Next, we then discuss our proposed a policy-based deep reinforcement learning model and the model training process that leads players to the correlated equilibrium in Section \ref{policy_based_deep_reinforcement_learning_neural_network} and Section \ref{model_training}, respectively. The second part is to estimate the payoff vectors of others via our proposed mathematical model that involves the idea of the force of tension in Section \ref{opponent_payoff_estimation}. In Section \ref{process_overview}, we summarize the entire process along with the use of our proposed models. 

\subsection{Process Overview}\label{process_overview}
In our system, we start with player $p_{main}$'s point of view. This means that at this point, we only know the information of player $p_{main}$ such as player $p_{main}$'s payoff vector $V_{main}$ and policy ${\pi}_{main}$. The process flowchart is shown in Fig. \ref{fig:process_flowchart} and outlined as followed.


We first find the set of players $\mathcal{P}_{against} \in P$ who obtain more than one Nash equilibrium when interacting with player $p_{main}$. We will say the player in set $\mathcal{P}_{against}$ are against to main player $p_{main}$ and the rest of the players are cooperating with player $p_{main}$. The reason we say the players in $\mathcal{P}_{against}$ are against to player $p_{main}$ is due to the payoff vectors of those players are monotonously increasing in the opposite direction as main player $p_{main}$. The detail on the monotonously increasing property will be discussed in Section \ref{opponent_payoff_estimation}.

Once we have set $\mathcal{P}_{against}$, we will let players $p_{main}$ and $p_t \in \mathcal{P}_{against}$ to interact with the system. During the interaction, both players will have to try their best to cooperate but also not to reveal too much information to others. This means each player not only has to discover the environment but also analyze the observed public information during the interaction. The public signals in our system model are the players' state after each action they performed. This information only tells a player what the other player's preference for each of the decision set $D_h$ but does not reveal the actual payoff vector of others. Some may argue that revealing the player's state gives too much information to the opponents. However, this is not true. Take the mixed strategy for example, although no player will know what the probabilities are been assigned to each pure strategy by the opponents, after certain rounds of game played, the player can base on the statistics to determine those probabilities. Therefore, revealing the players' states as the public signals is valid. Along with player's own information, each player will learn from their experience through their own deep neural networks which will be discussed in Section \ref{policy_based_deep_reinforcement_learning_neural_network}. Our proposed policy-based deep reinforcement learning model will learn the joint distribution between two players and lead them to the state that both players agreed on while still obtaining a certain amount of reward.

As for the termination of the interaction, since no player knows the payoff vector of the other player, we cannot terminate the interaction process based on a certain expect rewards of a player nor can we set a distribution goal to indicate the players have reached the correlated equilibrium distribution with an expect reward from both of the players. Hence, the only termination condition we can set is the number of actions the player can perform during the interaction. The number of actions has a restriction which will be discussed in Section \ref{data_collection} along with the interaction process and data collection process. After a certain amount of interactions, we can see that no matter how long the players interact with the environment, they will always be ending up in the same state. This means that the learning has been completed and they have reached state $s_{ce}$ where the probability distribution of decision set $\mathbb{D}$ satisfied the correlated equilibrium distribution definition and also maximize the joint rewards of both of the players based on this distribution.

Once we get the estimated correlated equilibrium distribution state $s_{ce}$, we can now go to the second part of our proposed method to estimate payoff vector ${\hat v}_t$ of player $p_t$. In order to estimate the payoff vectors, we proposed a method that combined the idea of the rationality defined in the theorem of correlated equilibrium in game theory and the idea of the force of tension. First, due to the fact that correlated equilibrium distribution is calculated with the rationality conditions regarding the payoff vectors of both of the players, when we calculate payoff vector $V_t$, the constraints still need to fulfill the rationality conditions. Next, we can treat the probabilities in the probability distribution as a type of preference for the decisions for each of the players. Based on the preference, we can determine the tension on a decision $D_h$ between each player. With these constraints, we can estimate payoff vector $V_t$ by solving a linear equation to maximize the reward of player $p_t$.

Once we estimate the payoff vector of player $p_t$, we will repeat this process until we go over each player in set $\mathcal{P}_{against}$ and estimated the payoff vector for all players in set $\mathcal{P}_{against}$. As mentioned before, the player who is not in the set of $\mathcal{P}_{against}$ is considered as the player whose cooperating with main player $p_{mian}$. This means these players who cooperate with player $p_{main}$ will most likely be against to player $p_t \in \mathcal{P}_{against}$. Hence, if we can set either one of the players in $\mathcal{P}_{against}$ as main player $p_{main}$, we can find the players who are against the new main player $p_{main}$ and repeat the process until we get the payoff vector for all players. Keep in mind that the termination condition can be triggered before each player interacts with each other players. Hence, this means some players will not interact with some other players at all. However, we can still compute the correlated equilibrium among those players simply by the correlated equilibrium theorem that we discussed in Section \ref{nash_equilibrium_and_correlated_equilibrium}.

\subsection{Data Collection}\label{data_collection}

We let player $p_l \in \left\{ {p_{main},p_t} \right\}$ interact with the environment for $M$ rounds independently. In the start of each round, player $p_l$ will always starts at the same state ${s_{def}}\in {\mathbb{S}}$. From this state, player $p_l$ will choose an action $a_j \in {\mathbb {A}}$ to perform according to $p_l$'s policy ${\pi}_l$. Once the action has been finished, player $p_l$ will reached to another state $s_k \in {\mathbb{S}}$. Player $p_l$ will than repeat the process of performing the next action and reaching to another state until player $p_l$ obtained $N$ states for each round $m$ for a total of $M$ rounds. The number of states, $N$, a player needs to obtain has to be an integer greater or equal to $\left\lceil {\vartheta ^{ - 1}} \right\rceil$. The reason is because since the player does not know the exact state which satisfied by the other player and each player must have the ability to discover the entire environment. Also, since each of the probability in a state $s_k$ is within the range of $\left[{0,1}\right]$, a minimum $\left\lceil {\vartheta ^{ - 1}} \right\rceil -1$ actions will guarantee the player has the ability to reach any state in the environment.

Back to the data collection process, player $p_l$ will record the number of action $a_k$, i.e., $k$, that performed at the $n$-th action in round $m$ in $c_{l,m,n}$ and the corresponding state in $s_{l,m,n}$ along the process. Here, the choice $c_{l,m,n}$ is an one-hot encode data. For example, if there are five actions in the action set $\mathbb{A}$ and the player chose action $a_2$ at the $n$-th action in round $m$, than the choice $c_{l,m,n}=01000$. Once $M$ rounds of interactions have been completed, player $p_l$ will have a set of actions' number $C_l$, i.e.,

\begin{equation}\label{data_collect_action_set}
    {C_l} = \left[ {\begin{array}{*{20}{c}}
        {{c_{l,1,1}}}& \cdots &{{c_{l,1,N-1}}}\\
        \vdots & \ddots & \vdots \\
        {{c_{l,M,1}}}& \cdots &{{c_{l,M,N-1}}}
        \end{array}} \right]
\end{equation}
and a set of corresponding states $S_l$, i.e.,

\begin{equation}\label{data_collect_state_set}
    {S_l} = \left[ {\begin{array}{*{20}{c}}
        {{s_{l,1,1}}}& \cdots &{{s_{l,1,N}}}\\
        \vdots & \ddots & \vdots \\
        {{s_{l,M,1}}}& \cdots &{{s_{l,M,N}}}
        \end{array}} \right].
\end{equation}
where $s_{l,m,1}=s_{def}$.

\begin{figure*}[!t]
   \centering
   \includegraphics[width=0.85\textwidth, height=7cm]{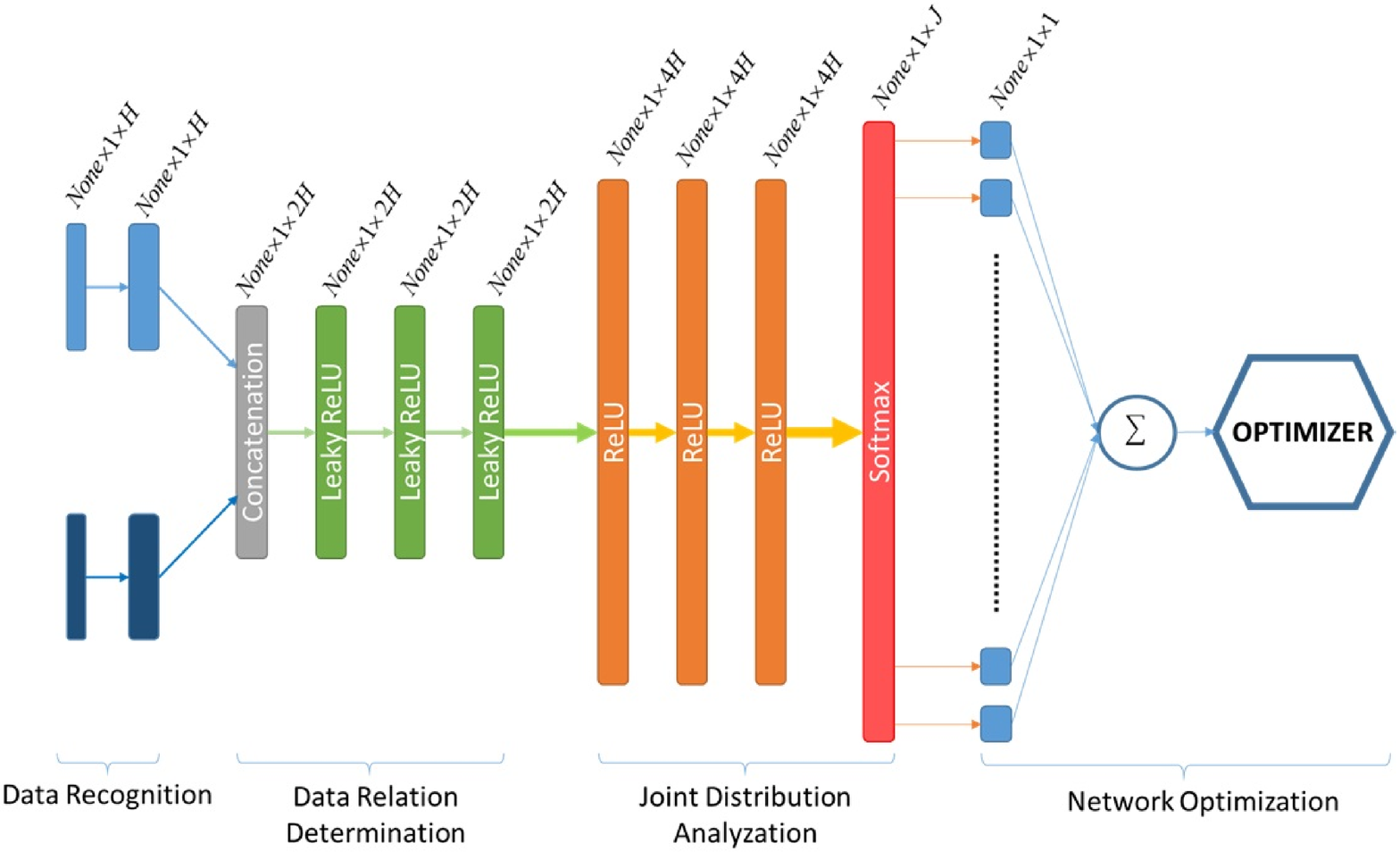}
   \caption{Policy-Based Deep Reinforcement Learning Neural Network Structure} 
   \label{fig:deep_neural_network}
\end{figure*}

Once player $p_l$ has finished interacting with the environment for $M$ rounds, we will need to calculate the rewards that player $p_l$ gained in each state in $S_l$. However, we cannot just calculate the reward for player $p_l$ based only on $S_l$ since for each action that player $p_l$ done is depends not only on player $p_l$'s policy ${\pi}_l$ but also depends on the action of the other player. Hence, when we calculate the reward gained in state $s_{l,m,n}$, we need to consider the $n$'s states in round $m$ from the other player where $1 \le m \le M$ and $1 \le n \le N$. The relationship between both of the $n$'s states in round $m$ from each of the players is that those states are equally important. Hence, the final state ${\hat s}_{m,n}$ for both players will be at the middle point of both of the $n$'s states in round $m$ from both players. Thus, in order to do determine the state ${\hat s}_{m,n}$, we need to gather both of the state sets from each player and averaging them element-wise to form the average state set $S_{avg}$, i.e.,

\begin{equation}\label{data_collect_state_avg_set}
    {S_{avg}} = \left[ {\begin{array}{*{20}{c}}
        {{{\hat s}_{1,1}}}& \cdots &{{{\hat s}_{1,N}}}\\
         \vdots & \ddots & \vdots \\
        {{{\hat s}_{M,1}}}& \cdots &{{{\hat s}_{M,N}}}
    \end{array}} \right]
\end{equation}
where

\begin{equation}\label{data_collect_state_avg}
    {{\hat s}_{m,n}} = \frac{1}{2}\sum\nolimits_{l = 1}^2 {{s_{l,m,n}}}.
\end{equation}

With the calculated average state set, we can then calculate the reward $r_{l,m,n}$ that player $p_l$ gained in the each of the state in the average state set $S_{avg}$ by computing the dot product of the average state ${\hat s}_{m,n}$ and the transport of the player's payoffs vector $V_l$, i.e.,
\begin{equation}\label{data_collect_reward }
    {r_{l,m,n}} = {\hat s}_{m,n} \cdot {V_l^T}.
\end{equation}

\subsection{Policy-Based Deep Reinforcement Learning Neural Network}\label{policy_based_deep_reinforcement_learning_neural_network}
The structure of our proposed policy-based deep reinforcement learning neural network (Fig. \ref{fig:deep_neural_network}) has two input layers where the input data for the first input layer will always be a set of states $s_{l,m,n}$ and the input data for the second input layer will be a set of states $s_{l,m,n-1}$ for $2 \le n \in N$. With these two input layers, the neural network can determine which direction the player is heading to. However, before determining the relationship between states $s_{l,m,n}$ and $s_{l,m,n-1}$, the network has to understand and recognize the information delivered by the probability distributions in each of the states. Therefore, right after each of the input layers, there is a hidden layer to analyze the input data. After the input analyzation layers, we concatenate the output of those two hidden layers with respect to the second axis of the data.

Followed by the concatenation layer are three sequential fully connected hidden layers. These three hidden layers are for the neural network to determine the relationship between the two input data. Since the difference between the two input data might outcome some negative values, the activation function for these three hidden layers should allow negative value to pass through. Moreover, as mentioned at the beginning of Section \ref{proposed_method}, the neural network works similarly as a classifier. This means the output values should only be positive numbers. Hence, the activation function for these three hidden layers are set to the leaky rectified linear unit (Leaky ReLU), i.e.,

\begin{equation}\label{dlm_leakyReLU}
f\left( x \right) = \left\{ {\begin{array}{*{20}{l}}
{0.2x},&{\text{for $x < 0$,}}\\
x,&{\text{for $x \ge 0$.}}
\end{array}} \right.
\end{equation}
This allowed certain negative values to pass through and also allowed the neural network to focus more on the positive values to match the output values.

What follows next is also another three sequential fully connected hidden layers. However, the difference is that the number of units in each of these hidden layers is twice the size of the layer in the previous three sequential hidden layers. Also, the activation functions in these three hidden layers are rectified linear unit (ReLU). The main purpose of these three hidden layers is to learn the joint distribution among all of the players in order for the deep neural network to determine the optimal action to perform.

Finally, the output layer is a fully connected layer with $J$ units which is same as the number of actions in the action set $\mathbb{A}$. The activation function in the output layer is the softmax activation function, i.e.,
\begin{equation}\label{dlm_softmax}
f{\left( x \right)_i} = \frac{e^{x_i}}{\sum\nolimits_{j = 1}^J {e^{x_j}}}
\end{equation}
for $i=1,\dots,J$. The purpose of the softmax activation function force the elements in the output vector to be within the range $\left( {0,1} \right)$ and the $L^1$-norm of the output vector is equal to 1. In other words, the output vector represents the probability distribution of the possible actions based on the input states.

\subsection{Model Training}\label{model_training}
We have discussed the structure of our proposed policy-based deep reinforcement learning neural network in the previous subsection where the output of the network is the probability distribution of the actions. Now we are going to discuss how we update the weights in the neural network. First of all, the output of the deep learning neural network works similarly as most of the multi-class classification but not exactly the same. In most of the multi-class classification, the loss function will be feed in a batch of output data along with the corresponding one-hot encoded labels. After each training via an optimizer, the probability for the correct predict label will be increased. There is nothing wrong with this process. However, most of the time, even if the input data only contains one category, some probabilities for the incorrectly predicted labels will also be slightly increased. As mentioned before in Section \ref{system_model}, the way we choose action $a_k$ is not based on the arguments of the maxima among the output vector but instead the probability distribution of the actions given by the output vector. Hence, when we increase the probability of one of the labels, we need to make sure the other probabilities do not increase. Therefore, the way we achieved this property is to calculate the loss for each of the probabilities. Here, we use the logarithmic loss for our loss function and scale the value by a weight. The absolute value of the weight indicates the change rate of the probability and the sign of the weight indicates whether should the probability move tower or away from the target for the positive and negative sign, respectively. The best and valid value to present as the weight for the loss function in our system is the reward $r_{l,m,n}$ that we have mentioned in Section \ref{data_collection}. However, we will need to perform post-processing on the reward data in order for our model to efficiently learn the joint distribution between the players.

In the post-processing procedure, we first multiply the reward $r_{l,m,n}$ by a discount factor $\gamma$ to the power of $N-n$ and get ${\bar r}_{l,m,n}$, i.e.,

\begin{equation}\label{data_collect_reward_gam}
{{\bar r}_{l,m,n}} = {{\gamma}^{\left( {N - n} \right)}}{r_{l,m,n}}
\end{equation}
where ${\gamma} \in \left[ {0,1} \right]$. This is due to the different importance of each action that the player performed during the interaction. Action $a_{l,m,N-1}$ will have more influence on which state player $p_l$ will be ended up on than the action $a_{l,m,N-2}$, and so on and so forth. In other words, we can consider actions $a_{l,m,n}$ for $n = 1,2, \ldots ,N$ is a time series data set since action $a_{l,m,n}$ will always be performed after action $ a_{l,m,n-1}$. In other words, $a_{l,m,N-1}$ will be the most recent action that player $p_l$ performed in round $m$. Based on the property of time series data, longer time horizons have much more variance as they include more irrelevant information, while short time horizons are biased towards only short-term gains. Hence, we need a discount factor to reduce the variance in the data set. With the calculated rewards ${\bar r}_{l,m,n}$, we get another set $\bar R$.

Now, the discount factor ${\gamma}$ has taken care of the variance caused by the long-term gains in the data set. There is another concern from the player's policy ${\pi}_l$ itself. As mentioned before in Section \ref{system_model}, player $p_l$ will sample an action according to the probability distribution given by their policy ${\pi}_i$ for the state that the player is currently in. This means that there is a chance that some actions will never or rarely be chosen based on the probability distribution. The problem with using the rewards from $\bar R$ to update the weights of the deep learning neural network is that the deep learning neural network will only be increasing the probabilities for sampled actions since reward ${\bar r}_{l,m,n}$ is always positive. Hence, the probabilities for those actions that have not been sampled or rarely sampled will be decreased throughout the training process. In order to overcome this problem, we will subtract the reward by a bias. Here, the bias for the reward is designed to be the mean ${\mu}_{l,n}$ of the rewards of the $n$-th observed state in the $M$ rounds of interaction. We also divided the reward by the standard deviation ${\sigma}_{l,n}$ of the rewards of the $n$-th observed state in the $M$ rounds of interaction. This will normalize the data for stability purpose during the training process. The post-processed reward ${\hat r}_{l,m,n} \in {\hat R}$ is expression as the following,
\begin{equation}\label{data_collect_reward_final}
{{\hat r}_{l,m,n}} = \frac{{{{\bar r}_{l,m,n}} - {\mu _{l,m}}}}{{{\sigma _{l,m}}}}
\end{equation}
where
\begin{equation}\label{data_collect_sigma}
{\sigma _{l,n}} = \sqrt {\frac{1}{M}\sum\nolimits_{m = 1}^M {{{\left( {{{\bar r}_{l,m,n}} - {\mu _{l,n}}} \right)}^2}}}
\end{equation}
and
\begin{equation}\label{data_collect_mu}
{\mu _{l,n}} = \frac{1}{M}\sum\nolimits_{m = 1}^M {{{\bar r}_{l,m,n}}}.
\end{equation}

After calculating the loss for each of the probabilities, we will sum up the losses and feed into the optimizer. In our proposed deep learning neural network, we applied Adam optimization with a learning rate of 0.001 to update the weights. We will keep repeating the process of data collection and model training until the player's last state in each round become stable before going to the next process of estimating the payoff vector of the other players.

\subsection{Opponent Payoff Estimation}\label{opponent_payoff_estimation}
Once we have determined the correlated equilibrium state $s_{ce}$, the next step will be estimating the payoff vectors for the other player by using the properties from both the decision choosing rationality in game theory and the force of tension. In addition, starting from this section, each player will do the calculation on their own. This means there will be no more information sharing among the players. Keep in mind that the payoff vectors estimation process works for every player in the system. However, as mentioned before, we will be discussing the process in player $p_{main}$'s point of view. Therefore, we only know the information about payoff vector $V_{main}$ and the probability distribution that is the correlated equilibrium state $s_{ce}$ at this point.

First, the correlated equilibrium distribution is based on the rational decision choosing by both of the players. The optimal probability distribution $\mathbb{P}$ for both of the players is the probability distribution that will maximize the joint reward of both of the players instead maximize their own reward. Therefore, under the circumstance where we know the exact payoff vectors for both of the players, the objective function $O_{old}\left({P} \right)$ for finding the optimal probability distribution $\mathbb{P}$ is expressed as follows,
\begin{equation}\label{obj_p_ce}
    \begin{aligned}
        \mathop {\max}\limits_{P} \quad & O_{old}\left({P} \right) = P \cdot {\left( {{V_{main}}} \right)^T} + P \cdot {\left( {{V_t}} \right)^T}
    \end{aligned}
\end{equation}
subject to
\begin{equation}
\begin{aligned}
  {\rho _h} \in P \ge 0,\\
  \sum\limits_{h = 1}^H {{\rho _h}} =1,\\
  \text{constraint (\ref{ce_def}).}
 \end{aligned}
\end{equation}

Now we want to reverse the calculation to estimate the payoff vectors $V_t $ when given the optimal probability distribution $\mathbb{P}$, the rationality conditions in (\ref{ce_def}) should still be fulfilled and the objective function is still be maximizing the reward summation of all players but based on the payoff vectors, i.e.,

\begin{equation}\label{obj_ce_p}
    \begin{aligned}
        \mathop {\max }\limits_{V_t} \quad & O_{new}\left({V_t} \right) = \mathbb{P} \cdot {\left( {{V_t}} \right)^T}.
    \end{aligned}
\end{equation}
Here, we can eliminate the reward of player $p_{main}$ since it is just a constant.

However, with (\ref{ce_def}) being the only constraint, the objective function will only be maximizing the payoff vectors as if each player's decision is independent with each other. As mentioned before, the constraint in (\ref{ce_def}) only gives a convex hull boundary on the probability distributions over the pure strategies. This means for each player, there will be at least one probability distribution that will benefit himself or herself but not necessary for others. Hence, even though the solution $\bar V_t$ will maximize the objective function $O_{new}\left({V_t} \right)$, the solution $\bar V_t$ will never equal to the actual payoff vectors with (\ref{ce_def}) being the only constraint. Moreover, if we try to solve the objective function $O_{old}\left({P} \right)$ with $\bar V_t$, the optimal solution will never be equal to $\mathbb{P}$. Hence, there need to be some other constraints that indicate the relationship between the players' payoff vectors. This is where the idea of the force of tension steps in.

\begin{figure}[t]
   \centering
   \includegraphics[width=0.85\columnwidth]{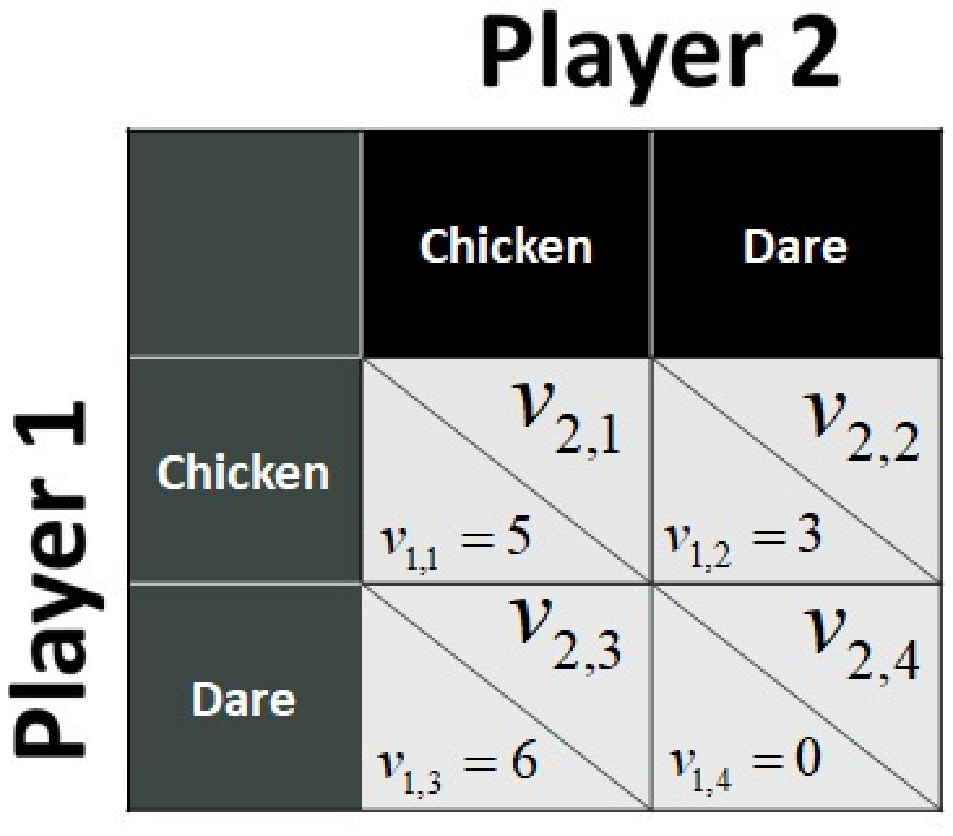}
   \caption{Game Setup for Payoff Vector Estimation} 
   \label{fig:payoff_estimate_illustration}
\end{figure}

As mentioned before in Section \ref{nash_equilibrium_and_correlated_equilibrium}, a distribution is a correlated equilibrium distribution if and only if it lies within the convex hull based on (\ref{ce_def}). In order to get the optimal probability distribution $\mathbb{P}$, the player needs to maximize the objective function $O_{old}\left({P} \right)$. In the force of tension point of view, we can consider optimal probability distribution $\mathbb{P}$ as an equilibrium point where the tensions for each decision set $D_h$ from both of the players have reached an equilibrium. Moreover, since the values in the optimal solution $\mathbb{P}$ are probabilities, we can treat those probabilities as the preferences of each decision set $D_h$ for the players. Based on this idea, we can list out the constraints regarding the relationship between the rewards in the payoff vector. There are a few types of constraints that we will be discussing in this section. For easier understanding, starting from here, we will use the game setup shown in Fig. \ref{fig:payoff_estimate_illustration} as the scenario where player $p_1$ and player $p_2$ are ``Player 1'' and ``Player 2'' in Fig. \ref{fig:payoff_estimate_illustration}, respectively. Keep in mind that we do not know the payoff vector $V_2$ at this time.

We first need to reorder the elements in the payoff vectors by sorting both of the payoff vectors and the probability distribution $\mathbb{P}$ based on the payoff vector $V_1$ in the ascending order. The payoff vector $V_2$ has to future rearrange the elements where the first element will be move to the end of the vector. After the reordering process, we get three new vectors ${V_1} \to {{\bar V}_1}$, ${V_2} \to {{\bar V}_2}$, $\mathbb{P} \to \bar {\mathbb{P}}$, and $\mathbb{D} \to \bar {\mathbb{D}}$, i.e.,
\begin{equation}
    {{\bar V}_1}  = \left\{ {{{\bar v}_{1,1}},{{\bar v}_{1,2}},{{\bar v}_{1,3}},{{\bar v}_{1,4}}} \right\} = \left\{ {{v_{1,4}},{v_{1,2}},{v_{1,1}},{v_{1,3}}} \right\},
\end{equation}
\begin{equation}
    {{\bar V}_2} = \left\{ {{{\bar v}_{2,1}},{{\bar v}_{2,2}},{{\bar v}_{2,3}},{{\bar v}_{2,4}}} \right\} = \left\{ {{v_{2,4}},{v_{2,2}},{v_{2,1}},{v_{2,3}}} \right\},
\end{equation}
\begin{equation}
    \bar {\mathbb{P}} = \left\{ {{{\bar \rho }_1},{{\bar \rho }_2},{{\bar \rho }_3},{{\bar \rho }_4}} \right\} = \left\{ {{\rho _4},{\rho _2},{\rho _1},{\rho _3}} \right\},
\end{equation}
and
\begin{equation}
    \bar {\mathbb{D}} = \left\{ {{{\bar D }_1},{{\bar D }_2},{{\bar D }_3},{{\bar D }_4}} \right\} = \left\{ {{D _4},{D _2},{D _1},{D _3}} \right\}.
\end{equation}

\begin{figure}[t]
   \centering
   \includegraphics[width=0.8\columnwidth]{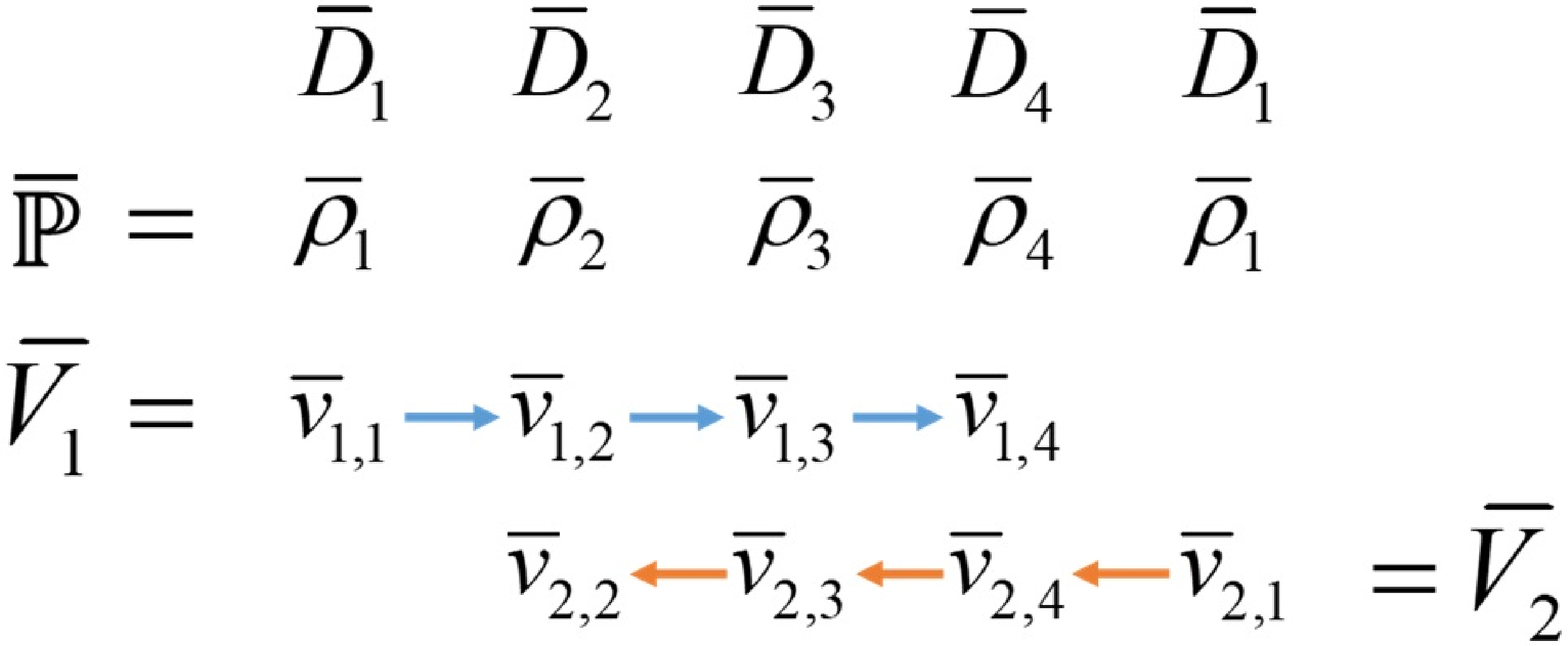}
   \caption{Payoff Vector Reorder} 
   \label{fig:payoff_reorder}
\end{figure}

The idea here is that players will try to move from the decision set with the lowest reward to the decision set with the highest reward where the direction for player $p_1$ is from the left to the right and the right to the left for player $p_2$. This setup is due to the monotonously increasing property for the convex set and function. The illustration is shown in Fig. \ref{fig:payoff_reorder} where we also showed the associated decision set for each reward. The arrow in the figure indicates the direction of the force that the player puts on the decision state. With this setup, we can start to discuss the types of conditions.

The first constraint type is regarding one outgoing force for each decision set from each player. We look at the first decision set $D_h$ where the corresponding rewards for both player $p_1$ and player $p_2$ have an outgoing arrow. We than subtract the force from player $p_1$ by the force from player $p_2$ to get the net force ${\vec f}_h$, i.e.,
\begin{equation}\label{diff_force_r}
    {\vec f}_h = {\bar \rho _{h+1}}\left( {{\bar v_{1,h+1}} - {\bar v_{1,h}}} \right) - {\bar \rho _{h-1}}\left( {{\bar v_{2,h-1}} - {\bar v_{2,h}}} \right)
\end{equation}
We can than list a constraint based on ${\delta}_{{\vec f}_h}$ where
\begin{equation}\label{diff_constraints_fr}
{\delta}_{{\vec f}_h} = \left\{ {\begin{array}{*{20}{l}}
{{{\vec f}_h} \ge 0,}&{{\text{if }}{\bar \rho _{h - 1}} < {\bar \rho _h} \le {\bar \rho _{h + 1}},}\\
{{{\vec f}_h} = 0,}&{{\text{if }}{\bar \rho _{h - 1}} \le {\bar \rho _h}{\text{ and }}{\bar \rho _{h + 1}} \le {\bar \rho _h},}\\
{{{\vec f}_h} \le 0,}&{{\text{if }}{\bar \rho _{h - 1}} > {\bar \rho _h} \ge {\bar \rho _{h + 1}}.}
\end{array}} \right.
\end{equation}
The reason for different signs of outgoing net force ${\vec f}_h$ is due to the preference according to probability $\bar{\rho}_h$. Reward $\bar{v}_h$ with higher probability $\bar{\rho}_h$ means it has a higher preference to the player which also means that the player will willing to move to from a reward with lower preference. Take reward $\bar{v}_{1,3}$ for example. If probability $\bar{\rho}_4$ is greater than probabilities $\bar{\rho}_3$ and $\bar{\rho}_2$, this means $p_1$ will be trying to move from $\bar{v}_{1,3}$ to $\bar{v}_{1,4}$. Since no player will be against to himself or herself, the only force that that will be against to player $p_1$ will came from player $p_2$ which is the force from $\bar{v}_{2,3}$ to $\bar{v}_{2,2}$. However, since the preference on reward $\bar{v}_{2,4}$ for player $p_2$ is also higher than reward $\bar{v}_{2,3}$ and $\bar{v}_{2,2}$, player $p_2$ will also be moving to reward $\bar{v}_{2,4}$ as well. Therefore, the force from reward $\bar{v}_{1,3}$ to reward $\bar{v}_{1,4}$ should be greater or equal to the force from player $p_2$ from reward $\bar{v}_{2,3}$ to reward $\bar{v}_{2,4}$. The same idea applied on other two signs in (\ref{diff_constraints_fr}).

The second constraint type is regarding one incoming force for each decision set from each player. We look at the first decision set $D_h$ where the corresponding rewards for both player $p_1$ and player $p_2$ has an incoming arrow. We than subtract the force from player $p_1$ by the force from player $p_2$ to get the difference ${{\mathord{\buildrel{\lower3pt\hbox{$\scriptscriptstyle\leftarrow$}} 
\over f} }_h}$, i.e.,
\begin{equation}\label{diff_force_l}
{{\mathord{\buildrel{\lower3pt\hbox{$\scriptscriptstyle\leftarrow$}} 
\over f} }_h} = {\bar \rho _h}\left( {{\bar v_{1,h}} - {\bar v_{1,h - 1}}} \right) - {\bar \rho _h}\left( {{\bar v_{2,h}} - {\bar v_{2,h - 1}}} \right)
\end{equation}
for

\begin{equation}\label{diff_force_l_cond}
{\bar v_{2,h - 1}} = \left\{ {\begin{array}{*{20}{l}}
{{{\bar v}_{2,1}},}&{{\text{if }}h = H,}\\
{{{\bar v}_{2,h - 1}},}&{{\text{otherwise}}}.
\end{array}} \right.
\end{equation}
We also can than list a constraint based on ${{\mathord{\buildrel{\lower3pt\hbox{$\scriptscriptstyle\leftarrow$}} \over f} }_h}$ where

\begin{equation}\label{diff_constraints_fl}
{\delta _{{{\mathord{\buildrel{\lower3pt\hbox{$\scriptscriptstyle\leftarrow$}} 
\over f} }_h}}} = \left\{ {\begin{array}{*{20}{l}}
{{{\mathord{\buildrel{\lower3pt\hbox{$\scriptscriptstyle\leftarrow$}} 
\over f} }_h} \ge 0,}&{}&{{\text{if }}{\bar \rho _{h - 1}} > {\bar \rho _h} \ge {\bar \rho _{h + 1}},}\\
{{{\mathord{\buildrel{\lower3pt\hbox{$\scriptscriptstyle\leftarrow$}} 
\over f} }_h} = 0,}&{}&{{\text{if }}{\bar \rho _{h - 1}} \le {\bar \rho _h}{\text{ and }}{\bar \rho _{h + 1}} \le {\bar \rho _h},}\\
{{{\mathord{\buildrel{\lower3pt\hbox{$\scriptscriptstyle\leftarrow$}} 
\over f} }_h} \le 0,}&{}&{{\text{if }}{\bar \rho _{h - 1}} < {\bar \rho _h} \le {\bar \rho _{h + 1}}.}
\end{array}} \right.
\end{equation}
for
\begin{equation}
{\bar \rho _{h + 1}} = \left\{ {\begin{array}{*{20}{l}}
{{{\bar \rho }_1},}&{{\text{if }}h = H,}\\
{{{\bar \rho }_{h - 1}},}&{{\text{otherwise}}.}
\end{array}} \right.
\end{equation}
The different signs of the incoming net force ${{\mathord{\buildrel{\lower3pt\hbox{$\scriptscriptstyle\leftarrow$}} \over f} }_h}$ is the same idea as we mentioned in constraint (\ref{diff_constraints_fr}).

We will repeat these two types of constraints with $L$ continuous outgoing force for the first type constraint and $L$ continuous incoming force for the second type constraint for $L = \left\{ {2,3, \ldots ,\left\lfloor {\frac{H}{2}} \right\rfloor } \right\}$. Finally, we will add few more constraints as follows,
\begin{equation}\label{v_ge}
v_{i,h} \ge 0,
\end{equation}

\begin{equation}\label{sum_v_ge}
\sum\limits_{h = 1}^H {{v_{i,h}}}  = 1,
\end{equation}
and the equality in the Nash equilibrium with the mixed strategy has to be established.

\subsection{Method Summary}\label{method_summary}
Now we have gone through the detail of each part of our proposed method. We can see that from our proposed method, all players can learn the joint distribution between each other with the policy-based deep reinforcement learning model and reach the correlated equilibrium. The correlated equilibrium probability distribution they reached allowed the player to obtain the maximum reward under a rational decision making in the game. With the correlated equilibrium probability distribution, the player can calculate the opponent's payoff vector based on our proposed mathematical model which involves the idea of the force of tension. The overall process is summarized in Algorithm \ref{overall_proposed_method} where we can see that not all players will interact with all other players. However, we still can compute the correlated equilibrium among those players who does not have interaction at all since we already have the payoff vectors of those players. This reduced the computation in respect of the entire system.

\begin{algorithm}[!t]
    \caption{Overall Proposed Method}
    \label{overall_proposed_method}
    \LinesNumbered
    \textbf{Definition:}\\
    $D_{i,j}$ is a set of all possible decision combination $d_{i,j,k}$ of all players in $P$ except player $p_i$ and $p_j$ where $i \ne j$;\\
    $NE_{i,j,k}$ be the set of the Nash equilibria with the mixed strategy between player $p_i$ and player $p_j$ when other players decision is fixed to $d_{i,j,k}$;\\
    Statement $f \left( p_i, {\pi}_i \right)$ indicates player $p_i$ will interact with the environment with policy ${\pi}_i$ based on $p_i$'s DNN to obtain $N$ states for $M$ rounds and return $C_i$ and $S_i$;\\
    Function $\Re \left( {\hat S,V_i} \right)$ calculates the post-processed reward set ${\hat R}_i$ for player $p_i$;\\
    \textbf{Initialization:}\\
    ${\Im}=\varnothing$\\
    \For{\rm{\textbf{each}} $p_i \in P$}{
        \For{\rm{\textbf{each}} $p_j \in P$ \rm{\textbf{AND}} $i \ne j$}{
            \For{\rm{\textbf{each}} $d_{i,j,k} \in D_{i,j}$}{
                \If{$\left\{ p_j, p_i, d_{j,i,k} \right\} \notin {\Im}$}{
                    ${\Im}$ append $\left\{ p_i, p_j, d_{i,j,k} \right\}$;\\
                }
            }
        }
    }
    \While{$\Im \ne \emptyset$}{
        \For{$\left\{ p_i, p_j, d_{i,j,k} \right\}$ \rm{in} ${\Im}$}{
            Other players' decision set to $d_{i,j,k}$;\\
            \If{$V_i$ \rm{\textbf{OR}} $V_j$ \rm{is known}}{
                \If{\rm{size}($NE_{main,i,d_{i,j,k}}$) $\ge 2$}{
                    \eIf{$V_i$ \rm{is known}}{
                        $p_m = p_i$ and $p_a = p_j$;\\
                    }{
                        $p_m = p_j$ and $p_a = p_i$;\\
                    }
                    ${\tilde {\mathbb{P}}}=\varnothing$\\
                    \While{$\tilde {\mathbb{P}}$ \rm{not stable}}{
                        $C_m, S_m=f\left( p_m, {\pi}_m \right)$;\\
                        $C_a, S_a=f\left( p_a, {\pi}_a \right)$;\\
                        $\hat S = 0.5 \left( {S_m+S_a} \right)$;\\
                        ${\hat R}_m=\Re \left( {\hat S,V_m}\right)$;\\
                        ${\hat R}_a=\Re \left( {\hat S,V_a}\right)$;\\
                        Update ${\pi}_m$ with $C_m$, $S_m$, and ${\hat R}_m$;\\
                        Update ${\pi}_i$ with $C_a$, $S_a$, and ${\hat R}_a$;\\
                        $\tilde {\mathbb{P}}={\hat S}\left[ -1 \right]$;\\
                    }
                    Set proposed constraints;\\
                    Maximize $O\left({{\tilde V}_a} \right)$ with respect to ${\tilde V}_a$;\\
                }
                Remove $\left\{ p_i, p_j, d_{i,j,k} \right\}$ from ${\Im}$;\\
            }
        }
    }
    Compute correlated equilibrium among players with $\mathbb{V}$;
\end{algorithm}

\section{Performance Evaluation}\label{performance_evaluation}
\subsection{Simulation Setup}\label{simulation_setup}

\begin{figure}[t]
   \centering
   \includegraphics[width=0.85\columnwidth]{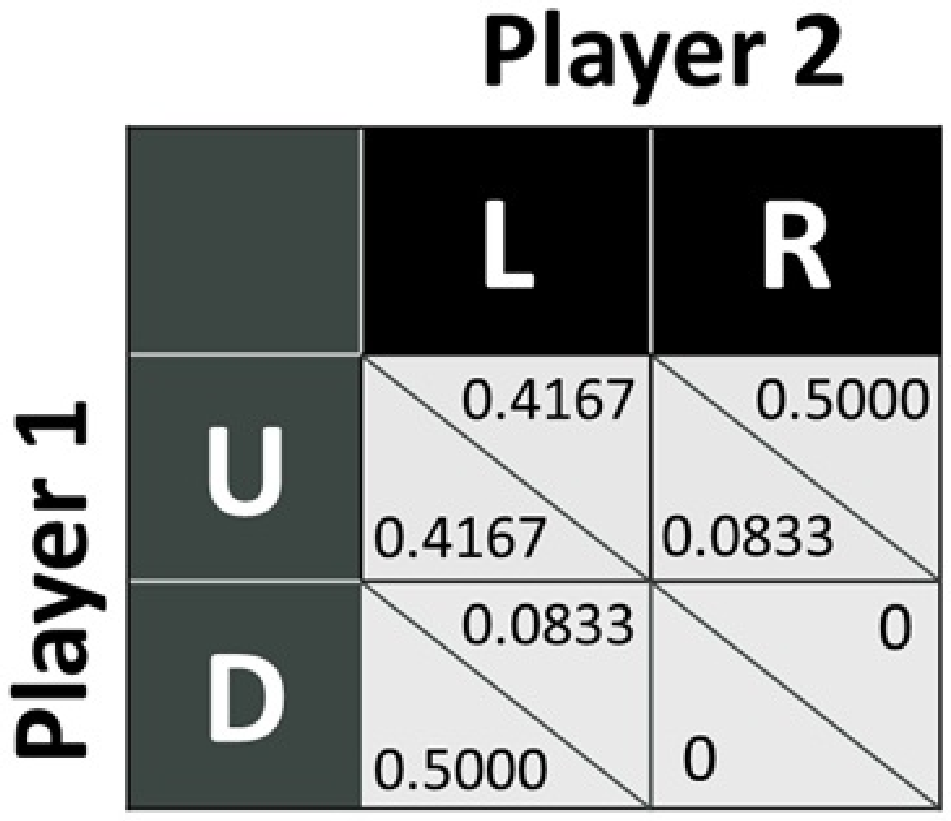}
   \caption{Simulation Setup} 
   \label{fig:simulate_setup_matrix}
\end{figure}

In our simulation, the game environment is setup based on two-person game where the players are player $p_1$ and player $p_2$. The decisions available to player $p_1$ are ``U'' and ``D'' where the decisions available to player $p_2$ are ``L'' and ``R''. Therefore, the decision sets $D_h$ for $h=\left\{ 1,2,3,4 \right\}$ are (U,L), (U,R), (D,L), and (D,R), respectively. The payoff vectors $V_1 = \left\{ 0.3571, 0.4286, 0.2143,0 \right\}$ and $V_2 = \left\{ 0.3571,0.2143,0.4286,0 \right\}$. The illustration of the setup is shown in Fig. \ref{fig:simulate_setup_matrix}. The step size $\vartheta $ is set to 0.005. Therefore, the action set $A$ is the permutation set among the elements in each of the three sets $\left\{ -0.005,0,0.005 \right\}$, $\left\{ -0.005,0,0.005 \right\}$, and $\left\{ -0.005,0,0.005 \right\}$ which has total of 27 actions. The player will interact with the environment for $M=40$ rounds and will perform $N=200$ actions in each round.

\begin{figure}[t]
   \centering
   \includegraphics[width=\columnwidth]{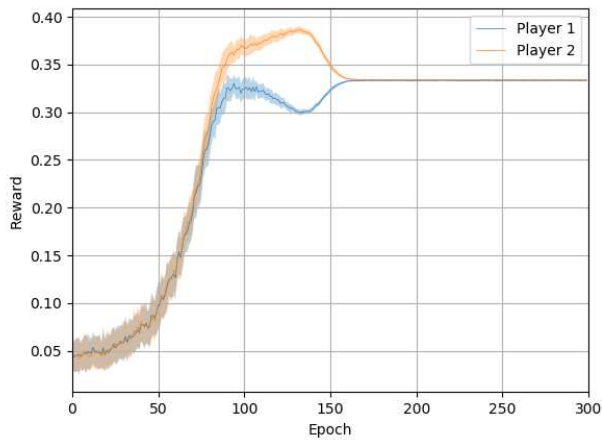}
   \caption{Average Reward Gained in Each Epoch} 
   \label{fig:avg_reward_epoch}
\end{figure}

\subsection{Numerical Results}\label{numerical_results}
With the setup of our simulation, we let the player $p_1$ be the main player. This means, at this point, we only know the payoff vector $V_1$ but not the payoff vector $V_2$. However, in order to illustrate the relation between both players, we will also show the data from player $p_2$ in the later on figures.

\begin{figure}[t]
   \centering
   \includegraphics[width=\columnwidth]{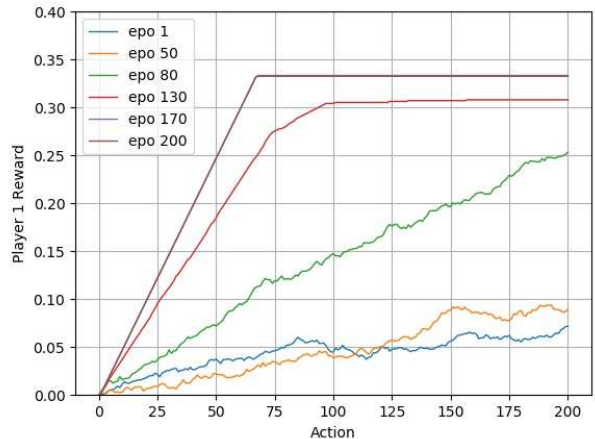}
   \caption{Player 1 Reward Increasing After Learning} 
   \label{fig:epoch_reward_imporve_p1}
\end{figure}
\begin{figure}[t]
   \centering
   \includegraphics[width=\columnwidth]{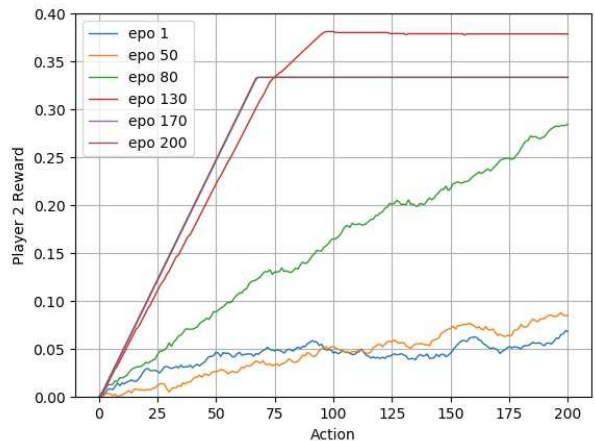}
   \caption{Player 2 Reward Increasing After Learning} 
   \label{fig:epoch_reward_imporve_p2}
\end{figure}
\begin{figure}[t]
   \centering
   \includegraphics[width=\columnwidth]{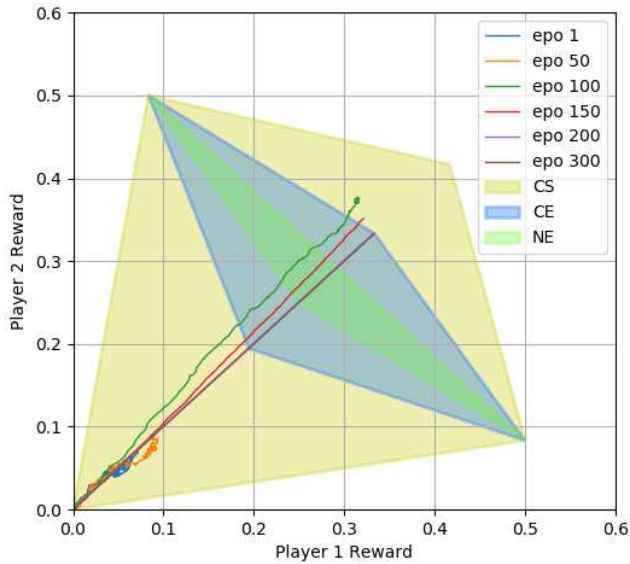}
   \caption{Track of Both Players After Learning} 
   \label{fig:epoch_reward_improve_comb}
\end{figure}

We first let two players interact with the environment for more than 300 epochs where the player interacts 40 rounds in each epoch. As we can see in Fig. \ref{fig:avg_reward_epoch}, both players are gaining small and unstable rewards at the beginning. As they interact with the environment with their own policy-based deep reinforcement learning neural network in each epoch, they learn from the environment and the opponent and receiving higher rewards as the number of epochs increases. In addition, as the more they learned, the faster they reach the higher reward. The results were shown in Fig. \ref{fig:epoch_reward_imporve_p1} and Fig. \ref{fig:epoch_reward_imporve_p2} for player $p_1$ and player $p_2$, respectively. However, around 80 epochs where they reach mixed Nash equilibrium, they started to diverge from each other. More specifically, player $p_2$ starts to pull player $p_1$ tower decision set ${D_2}= \left( \text{U,R} \right)$. Around epoch 130, player $p_1$ realizes that he or she must act to stop the reduction on his or her reward. Hence, player $p_1$ starts to pull back player $p_2$. Around epoch 170, both players reach an equilibrium point where the state at the end of each round of interaction was stabled at $\left\{\frac{1}{3}, \frac{1}{3},\frac{1}{3},0 \right\}$. In Fig. \ref{fig:epoch_reward_improve_comb}, we can see the track of the interaction from both players in different epochs where the areas of ``CS'', ``CE'', and ``NE'' represent the set of correlated strategies, the convex hull of the Nash equilibrium, and the convex hull of correlated equilibrium, respectively. We can see that as they observe the environment and learn the opponent's behavior, they achieve outside of the Nash equilibrium convex hull and become stable at the correlated equilibrium where maximizes both players' reward. Even though we simulate more than 500 epochs, the finial state remains stable after epoch 170. Hence, we only showed the result of up to 300 epochs in each figure.

Once the both players have reached a stable state after well trained, we use this state as the estimate correlated equilibrium probability distribution where $\tilde {\mathbb{P}} = \left\{\frac{1}{3}, \frac{1}{3},\frac{1}{3},0 \right\}$ to determine the estimate payoff table for player $p_2$. Moreover, if we calculate the correlated equilibrium with the objective function (\ref{obj_p_ce}), we can see that the solution $\mathbb{P}=\left\{\frac{1}{3}, \frac{1}{3},\frac{1}{3},0 \right\}$ is exactly the same as the probability distribution in state $s$. 

With the estimate correlated equilibrium probability distribution $\tilde {\mathbb{P}}$, we are able to list out the objective function and the constraints based on the steps in Section \ref{opponent_payoff_estimation}.
By solving this constrained linear multivariable function, we get the estimate payoff vector ${\tilde V_2} = \left\{ {0.4167,0.5000,0.0833,0} \right\}$ of player $p_2$. Compare with the actual payoff vector $V_2 = \left\{ {0.4167,0.5000,0.0833,0} \right\}$, we can see that there is no error between these two vectors. Hence, we have successfully computed the payoff vector for the other player.


In the beginning, the players can only apply the mixed strategy to reach the Nash equilibrium and obtain the reward 0.25 and 0.25 for player $p_1$ and player $p_2$, respectively. Now, with our proposed deep reinforcement learning model, both players can obtain a higher reward of 0.3333 and 0.3333 for player $p_1$ and player $p_2$, respectively, by reaching the estimated correlated equilibrium.



\begin{figure}[t]
   \centering
   \includegraphics[width=\columnwidth]{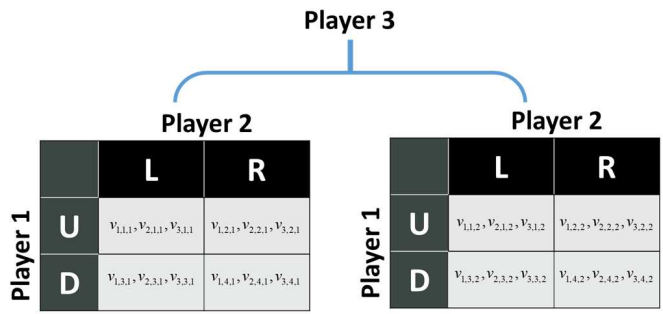}
   \caption{Three Players Game} 
   \label{fig:3players}
\end{figure}

Now, in the case that we have three players in the game as shown in Fig. \ref{fig:3players} where player three $p_3$ can choose left or right matrix they will be playing. With player $p_3$ chooses the left matrix, we will do the same thing as before where we let player $p_1$ interact with player $p_2$ and get the payoff vector $V_2$. However, we will also let either player $p_1$ interact with player $p_3$. By doing so, we can get the payoff vector $V_3$. By having the payoff vectors $V_2$ and $V_3$, we can compute the correlated equilibrium between player $p_2$ and player $p_3$ without them to interact with each other. The same thing for the case where player $p_3$ chooses the right matrix. This way, we reduced the computation of the interaction between player $p_2$ and player $p_3$ in the entire process.

\section{Discussion of Application Scenarios}\label{discussion}
The correlated equilibrium solution concept has been adopted in many different areas. However, most of them still relied on either a centralized system or information sharing between parties. In this section, we will be discussing the applications that involve the correlated equilibrium in wireless communication, smart grid, and resource allocation, where our proposed scheme can reduce signaling and improve the performance.

We first look at the applications in wireless communication where the most common wireless technologies use radio waves. With radio waves, the transmission distances can be as short as few centimeters such as NFC and as far as millions of miles for deep-space radio communications. However, there are some challenges in wireless communication as signal interference, data throughput, and more need to be solved in order to have a stable communication between devices. The author in \cite{6096766} proposed a distributed cooperation policy selection scheme for interference to perform subcarrier assignment for uplink multi-cell OFDMA systems by adopting the correlated equilibrium solution concept that achieves better performance by allowing each user to consider the joint distribution among users' actions to minimize the interference. Also, in \cite{8753494}, the author showed by using a game-theoretic learning algorithm which is based on correlation equilibrium in the problem of multi-user multichannel access in distributed high-frequency diversity communication networks can completely avoid interference and get optimal throughput. Moreover, it also guarantees fairness among all user equipment. Besides the concerns on the interference in wireless communication, the data throughput is also a main concern as well. In the work \cite{6882640}, the author showed a game-theoretic approach based on correlated equilibrium and regret-matching learning can provide significant gains in terms of average cell throughput in the Monte Carlo simulations of Long Term Evolution - Advanced like system. 

Next, the smart grid is an electrical grid that includes a variety of operation and energy measures including smart meters, smart appliances, renewable energy resources, and energy-efficient resources \cite{7454295}. Electronic power conditioning and control of the production and distribution of electricity are important aspects of the smart grid \cite{lee2013assessment}. In energy-aware ad hoc networks, energy efficiency is a crucial requirement. The authors in \cite{6392311} present a cooperative behavior control scheme based on the correlated equilibrium to reduce and balance energy consumption.

Finally, in resource allocation problem which arises in many application domains ranging from the social sciences to engineering \cite{4215753,6717015}, the objective is to allocate resources to different areas under some concerns such as energy consumption, fairness, and more. In \cite{6213824}, the authors proposed an energy-efficient resource allocation scheme by using the correlated equilibrium. Furthermore, the authors present a linear programming method and a distributed algorithm based on the regret matching procedure to implement the CE. With their proposed method, they can determine the desired resource allocation in an uplink orthogonal frequency division multiple access (OFDMA) system.

\section{Conclusion}\label{conclusion}

In this work, we have successfully overcome the issue regarding the public signal in the correlated equilibrium solution concept through our proposed deep reinforcement learning model. We can see from the numerical results that the model learned the joint distribution of all the players and reached the state of the correlated equilibrium probability distribution. Moreover, with the information from the player himself or herself and the correlated equilibrium probability distribution, achieved from the deep reinforcement learning model, we propose a mathematical model to estimate the payoff vector of the other player, which combines the concept of the rationality in game theory and the force of tension. Once we have the estimated payoff vectors of other players, we can compute the correlated equilibrium among the player and the other players who have not been interacted with each other. This paper combines the game theory with machine learning in the sense that the proposed machine learning learns what is the game player's payoff, instead of just categorizing the strategies of actions according to the current situation.

\ifCLASSOPTIONcaptionsoff
  \newpage
\fi



\bibliographystyle{IEEEtran}
\bibliography{reference}
\end{document}